\documentclass[aps,pre,showpacs,floatfix]{revtex4} 
\usepackage{epsfig} 
  
\begin{document}  
  
\def\r{{\rm r}}  
\def\dr{{\Delta{\rm r}}}   
\def\C{{\bf C}}  
 
\title{Hierarchical Structure in Healthy and  
Diseased Heart Rate Variability in Humans}  
\author{Emily S.C. Ching$^1$, D.C. Lin$^2$ and C. Zhang$^1$}  
\affiliation{ 
$^1$Department of Physics, The Chinese University of Hong Kong,  
Shatin, Hong Kong.\\  
$^2$Department of Mechanical and Industrial Engineering, 
Ryerson University, Toronto, Canada.}  
  
\begin{abstract}  
It is shown that the heart rate variability (HRV)   
in healthy and diseased humans possesses a hierarchical structure  
of the She-Leveque (SL) form. This structure, first found in  
measurements in turbulent fluid flows, implies further details  
in the HRV multifractal scaling. The potential of diagnosis is also  
discussed based on the characteristics derived from the SL hierarchy.  
\end{abstract}  
  
  
\maketitle  
  
The heart beat interval in humans is known to exhibit 
fluctuation which is referred to as heart rate
variability (HRV). Power spectrum analysis of the fluctuation 
revealed a $1/f$-like scaling \cite{ref3}. Recent studies indicated that
healthy human HRV exhibits even higher complexity which 
can be characterized by multifractal scaling \cite{ref1,ref2}. 
In contrast, HRV in the 
pathological state such as congestive heart failure
exhibits more monofractal-like scaling \cite{ref1}.
The change of the HRV 1/f law in congestive heart
failure is consistent with this result \cite{ref4}.
Such a multifractal-monofractal transition was also
reported in parasympathetic nervous system (PNS)
blockade experiment \cite{ref2}. Hence the
manifestation of multifractal HRV is indicative of the
proper autonomic regulation of the heart rate. Further
studies revealed that the multifractal HRV have
properties analogous to those found in fluid turbulence
\cite{ref5}. However, there is little understanding
beyond the phenomenological description of multifractal
HRV.
 
In this paper, we exploit further the 
analogy of HRV to fluid turbulence
and show the existence of a hierarchical 
structure in healthy and diseased HRV. This structure
allows us to model the multifractality of HRV and make
conjecture to the heart beat dynamics responsible for
the multifractal scaling. The hierarchy, first proposed
by She and Leveque (SL) to understand the statistical
properties of turbulent flows, provides a successful
framework to discuss and characterize the deviation
from Kolmogorov monofractal scaling in fluid turbulence
\cite{ref6}. When applied to study HRV, the SL
hierarchy provides a model structure which possesses
two advantages: (a) it simplifies the functional
description of the multiscaling by using a maximum of
only three parameters, and (b) it contains predictive
power for HRV scaling in physiological states related
to PNS withdrawal. One immediate implication is the
potential use of this notion in applications such as
diagnosis.
 
Let the beat-to-beat RR interval (RRi) be $\r(t)$, where 
$t$ is the beat number, and its increment be $\dr(\tau)= 
\r(t+\tau)-\r(t)$. The SL hierarchy implies, for a range 
of $\tau$, 
\begin{equation} 
\left[{S_{p+2}(\tau)\over S_{p+1}(\tau)}\right] =  
A_{p}\left[{S_{p+1}(\tau)\over S_p(\tau)}\right]^\beta  
\left[S^\infty(\tau)\right]^{1-\beta}. 
\label{1}  
\end{equation} 
Here $0 < \beta < 1$ is a parameter 
of the hierarchy, $A_p$, a function of $p$, $S_p(\tau)= 
\langle|\dr(\tau)|^p\rangle$, the $p$-th order moment of 
$|\dr(\tau)|$ denoted as the $p$-th order RRi structure 
function, $S^\infty(\tau)\equiv\lim_{p\to\infty}S_{p+1}( 
\tau)/S_p(\tau)$ and $\langle\cdot\rangle$ denotes 
statistical average. Since $S^\infty(\tau)$ is dominated 
by the statistics of large $\dr(\tau)$, it characterizes
the most intense fluctuation in HRV. Moreover, given the
empirical law $S_p(\tau)\sim\tau^{\zeta(p)}$ in HRV
\cite{ref5}, the hierarchy (\ref{1}) implies \cite{ref6} the scaling
model 
\begin{equation} 
\zeta(p) = h_0 p + C ( 1 - \beta^p)  
\label{2}  
\end{equation}  
where $h_0$ and $C$ are two other parameters of the hierarchy. 
It follows from (\ref{1}) and (\ref{2}) that $S^\infty(\tau)
\sim\tau^{h_0}$. A nonlinear functional dependence of $\zeta(p)$
on $p$ indicates multifractal scaling. Thus, the parameter
$\beta$ measures the degree of multifractality. In particular,
$\beta\to 1$ leads to monofractal scaling. In the multifractal
description of fluid turbulence, the parameter $C$ can be shown
to be the codimension of the most intense structure of the flow
\cite{ref6}. Since it is not yet possible to write down the
equation of motion for long-term cardiovascular dynamical system,
we assume a working definition for $C$ as the ``codimension
parameter" of the signal.
 
We follow the procedure developed in Ref. \cite{ref7} to check
whether the RRi data possess a SL hierarchical form. This
approach, based on the scaling property implied by the
hierarchy, the so-called generalized extended self-similarity
(GESS) in fluid turbulence \cite{ref8,ref9}, describes a
power-law relationship between the normalized structure
functions: 
\begin{equation}  
{S_p(\tau) \over [S_n(\tau)]^{p/n}} \sim   
\left\{{S_q(\tau) \over [S_n(\tau)]^{q/n}}\right\}^{\rho_n(p,q)} 
\label{5}  
\end{equation}  
In the case of SL hierarchy, the exponents $\rho_n(p,q)$ depends 
only on the model parameter $\beta$: 
\begin{equation}  
\rho_n(p,q) = {n(1-\beta^p)-p(1-\beta^n) \over  
n(1-\beta^q)-q(1-\beta^n)}. 
\label{6}  
\end{equation} 
It follows that 
\begin{equation}  
\Delta \rho_n(p+\delta p,q) = \beta^{\delta p} \Delta \rho_n(p,q)  
 - {\delta p(1-\beta^n)(1-\beta^{\delta p}) \over n(1-\beta^q)-q(1-\beta^n)}  
\label{7}  
\end{equation}  
where $\Delta \rho_n(p,q) \equiv \rho_n(p+\delta p,q)-\rho_n(p,q)$. 
One can then plot $\Delta \rho_n(p+\delta p,q)$ vs $\Delta
\rho_n(p,q)$ to check (\ref{7}) and hence the validity of the SL
hierarchy. We use several databases to perform the calculations (3)
and (5). The first database (DB1) contains 10 sets of daytime
ambulatory RRi recordings taken from healthy young adults \cite{ref5}.
The second database (DB2) contains 18 sets of daytime normal sinus
rhythm RRi data downloaded from public domain \cite{ref10}. We also
analyze RRi data from congestive heart failure patients (DBCHF) from
the same public domain \cite{ref10} to study the intrincity of the
hierarchy.
 
Except for a few cases where excessive ectopic beats in the RRi 
data complicates the calculation of $S_p(\tau)$, and were 
therefore discarded from the analysis, GESS is found in both 
healthy and CHF HRV. The exponent $\rho_n(p,q)$ is then 
estimated from (\ref{5}) and used in (\ref{7}) to calculate 
$\Delta\rho_n(p,q)$. Typical $\Delta \rho_n(p+\delta p,q)$ vs 
$\Delta \rho_n(p,q)$ plots are shown in Fig.~1. The observed 
linear trend implies (\ref{7}). Hence, SL hierarchy is 
compatible with the multifractal scaling in HRV. From such
plots, we estimate the value of $\beta$ simultaneously from
the slope and the intercept of the fitted straight lines. The
results are given in Fig.~2a. The $\beta$'s from healthy HRV
(DB1,DB2) cluster in the range [0.65,0.85] with those $\zeta 
(p)$ showing less curvature being characterized by larger 
$\beta$ values (Fig.~1). The $\beta$'s from DBCHF are generally
larger in values due to the monofractal-like scaling.
  
To gain insight of the hierarchy, She and Waymire (SW) arrived 
at the hierarchy (\ref{1}) using multiplicative random cascade 
\cite{ref11}. Their cascade consists of two dynamic components. 
One is the basic component that generates the singular 
dynamics over a continuum of scales. It can be shown that this 
dynamical component gives rise to the scaling term $h_0p$ in 
(\ref{2}). SW's cascade contains an extra component, which 
they called the ``defect dynamics" (DD), that modulates the 
singular structure through the multiplication of $\beta$ in 
discrete steps \cite{ref11}. It can be shown that this dynamical 
component contributes to the nonlinear term $C(1-\beta^p)$ in 
(\ref{2}). To apply SW's interpretation requires some 
clarification since continuous scale invariance does not exist 
in HRV, due to the fact that RRi fluctuation between heart 
beats cannot be defined. This suggests a dominant DD in the
generation of the multifractal scaling of HRV and a scaling
model with $h_0\sim 0$. Equivalently, this implies a hierarchy
with a $\tau$-independent $S^\infty$. Since $S^\infty$ cannot
be directly calculated, to verify such a model we first re-write
(\ref{1}) as  
\begin{equation}  
S_p(\tau) \sim [S^\infty(\tau)]^p\left\{S_q(\tau)  
\over S^\infty(\tau)^q\right\}^{\mu(p,q)}  
\label{8}  
\end{equation}  
where $\mu(p,q) \equiv (1-\beta^p)/(1-\beta^q)$ \cite{ref7}. We then 
form the quotient of (\ref{8}) at distinct values of $\tau$ and $\tau_0$, 
which, after some algebra, yields 
\begin{equation} 
\log_2\left[{S^\infty(\tau)  
\over S^\infty(\tau_0)}\right]  
={\log_2[S_p(\tau)/S_p(\tau_0)]  
-\mu(p,q)\log_2[S_q(\tau)/S_q(\tau_0)]\over  
p-q\mu(p,q)}  
\equiv F_{p,q}(\tau,\tau_0).
\label{9}  
\end{equation}  
Hence, $F_{p,q}(\tau,\tau_0)$ is independent of $p$ and $q$ and  
a $\tau$-independent $S^\infty(\tau)$ implies a ``constant"
$F_{p,q}(\tau,\tau_0)$ over a range of $\tau$ and $\tau_0$
values. Figure~3 shows $F_{p,q}(\tau,\tau_0)$ for healthy and
CHF HRV. It is seen that the condition $h_0\sim 0$ can be
statistically ascertained and that $\zeta(p) \sim C(1-\beta^p)$.
Given this, $C$ is obtained by averaging $\zeta(p)/(1-\beta^p)$
over a range of $p$ and its result has been shown in Fig.~2b. 
Moreover, $C$, as a function $\beta$, shows an increasing trend
as $\beta\to 1$ (Fig.~4). This functional relationship is
consistent with the observations $h_0\sim 0$ and that $\zeta(p)$
becomes almost proportional to $p$ as $\beta \to 1$. It can
also be inferred from the earlier experimental studies, as we now explain.

Recall that the fractal dimension $D(h)$ of the set with a local
scaling exponent $h$ is related to $\zeta(p)$ through a Legendre
transform:
\begin{equation}  
D(h) = {\rm min}_p [ph + d - \zeta(p)]   
\label{3}  
\end{equation}  
where $d$ is the dimension of the embedding space. From (\ref{2})
and with $h_0\approx 0$, $D(h)$ can be explicitly obtained as: 
\begin{equation}  
D(h) = d - C + \left[ 1+ \ln C + \ln (\ln 1/\beta) \over \ln(1/\beta)  
\right]h    
 - {h \ln h \over \ln(1/\beta)}  
\label{4}  
\end{equation}  
Let $h^*$ be the scaling exponent of the singular
set with largest dimension, i.e., $D(h^*)$ is the maximum. 
For HRV, $h^*$ was
found to increase its value from the multifractal-like
scaling in healthy state to the monofractal-like scaling
in the diseased and pathological states \cite{ref1,ref2}.
Using (\ref{4}), $h^*$ is derived explicitly as
\begin{equation}  
h^*=C\ln(1/\beta).  
\label{10}  
\end{equation}  
If $C$ is constant, $h^*$ decreases as monofractal scaling 
is approached ($\beta\to 1$), which contradicts what was 
observed \cite{ref1,ref2}. In order for $h^*$ to increase
as $\beta \to 1$, $C(\beta)$ must diverge as $1/\ln (1/
\beta)$. Indeed, we find that the dependence of $C$ on
$\beta$ can be well described by $0.2\beta/\ln(1/\beta)$
(Fig.~4). Thus we have the result $h^*\sim 0.2\beta$
with $h^*$ increasing with $\beta$ in accord with the
experimental observations \cite{ref1,ref2}.
Finally, the asymptotic behaviour of $C$ (as $\beta\to 1$)
presents a more favorable condition for diagnosis: i.e., the
more monofractal-like scaling falls into the range with
larger ``separation" of the $C$ values (see also Fig.~2b).  
  
In summary, we show that a hierarchical structure of the SL 
form exists in the healthy and diseased HRV. This property 
allows us to model the multifractal HRV in terms of only 
two parameters $C$ and $\beta$. Interestingly, $C$ and
$\beta$ are related by an empirical law captured in Fig.~4.
This finding is important for two reasons. First, the
empirical law appears universal and is capable of describing
both healthy and CHF data. Second, the divergence of $C$ as
$\beta\to 1$ implies potential in diagnosis using the
hierarchical structure. To find a model that is compatiable
with the current finding, we adopted SW's cascade which
leads to further implications beyond the phenomenological
description of multifractal HRV scaling. This model is
appealing in that it contains a modulating component (DD)
acting on the singular dynamics to ``tame" the
fluctuation (since $\beta<1$.) Its effect reminds us of
the function of feedback regulation in biological systems.
Hence, the SW's model provides a concrete example of how
(additive) feedback mechanism may be integrated
multiplicatively in a cascading structure to produce the
observed HRV phenomenology. Further experiments will be
needed to test and quantify these possibilities in more
detailed physiological terms.
  
\begin{center}  
{\bf Acknowledgment}  
\end{center}  
ESCC acknowledges the Hong Kong Research Grants Council (CUHK 
4286/00P) for support. D.C. Lin would like to thank the hospitality
of the Physics Department at the Chinese University of Hong Kong,
the C.N. Yang Visiting Fellowship and Natural Science and
Engineering Research Council of Canada for supports.

\newpage 
 
\begin{figure}
\centering
\includegraphics[width=.45\textwidth,angle=90]{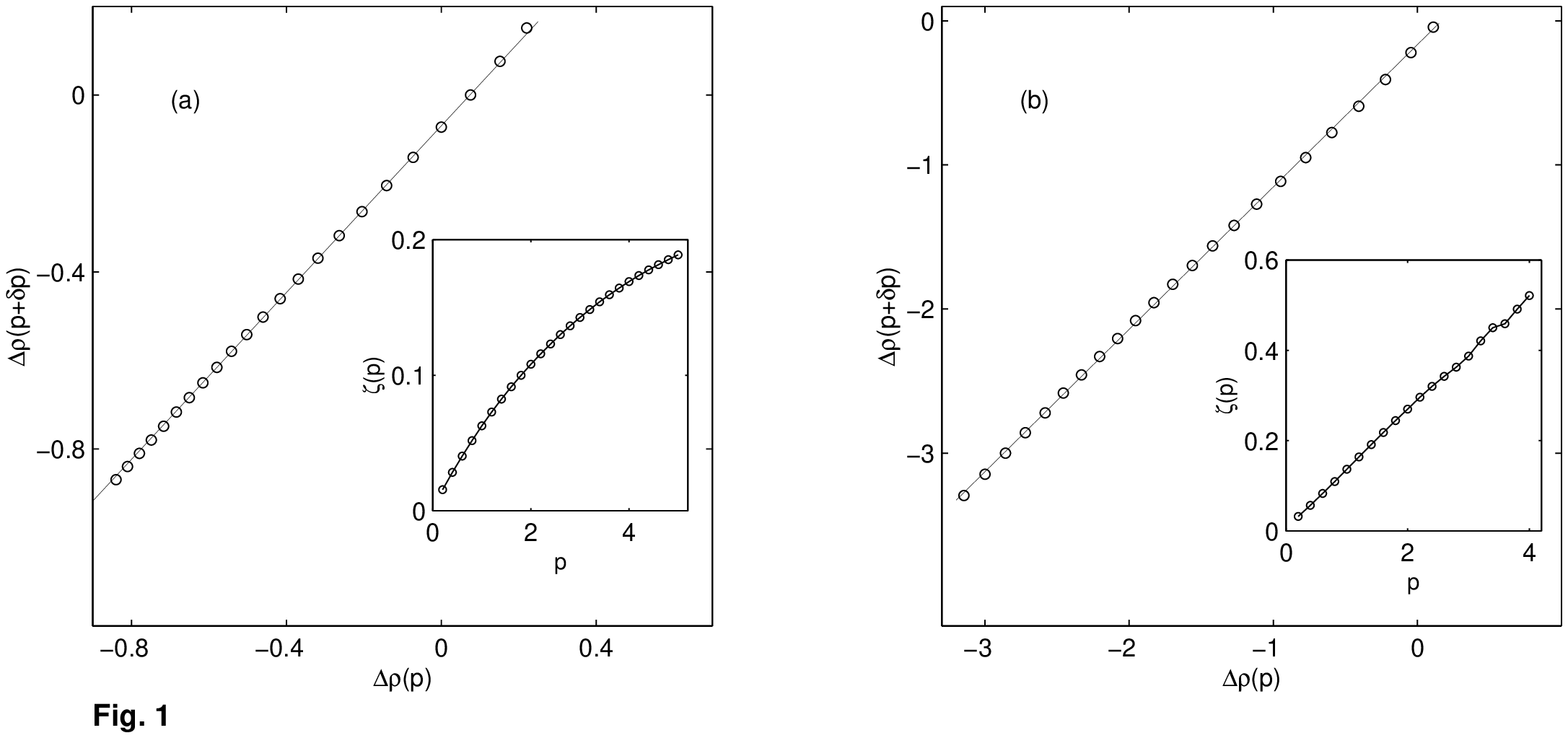}
\caption{
Typical $\Delta\rho_n(p+\delta p,q)$ vs.
$\Delta\rho_n(p,q)$ plots for (a) healthy HRV and (b) CHF
HRV with $\delta p=0.2$. We used $q=1$ and $n=2$ for
healthy HRV and $q=0.8$ and $n=1.2$ for CHF HRV. The
arrow direction indicates the increasing $p$ direction.
The estimated slopes for the shown cases are $\beta^{
\delta p}$ = 0.9422 and 0.9881, respectively. Hence,
$\beta\sim(0.9422)^5=0.7425$ and $(0.9881)^5\sim 0.9491$,
respectively. The insets show the corresponding
$\zeta(p)$. It is seen that a large $\beta$ in CHF HRV
implies a $\zeta(p)$ with less curvature, i.e.,
monofractal-like scaling.}
\end{figure}

\begin{figure}
\centering
\includegraphics[width=.95\textwidth]{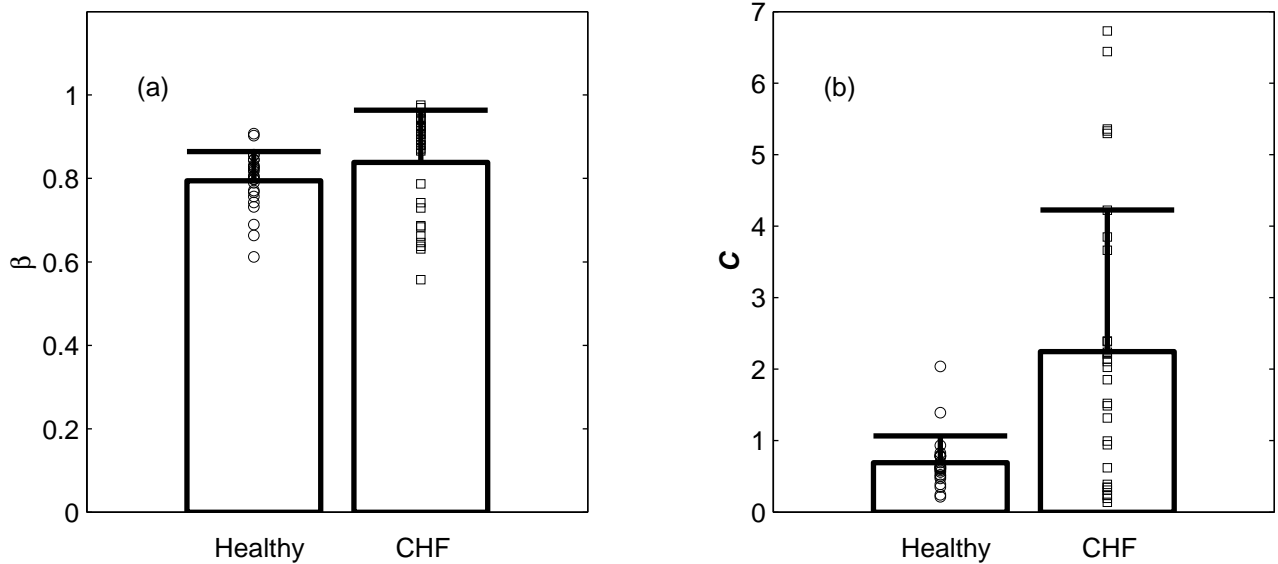}
\caption{Estimated values for model parameters (a) $\beta$ and
(b) $C$. Estimates from DB1, DB2 for 24 healthy subjects
are given in circles and those from DBCHF for
30 congestive heart failure subjects are given in squares.
Means and standard deviations in both cases are also shown by the
bar-chart.}
\end{figure}
  
\newpage

\begin{figure}
\centering
\includegraphics[width=.6\textwidth,angle=90]{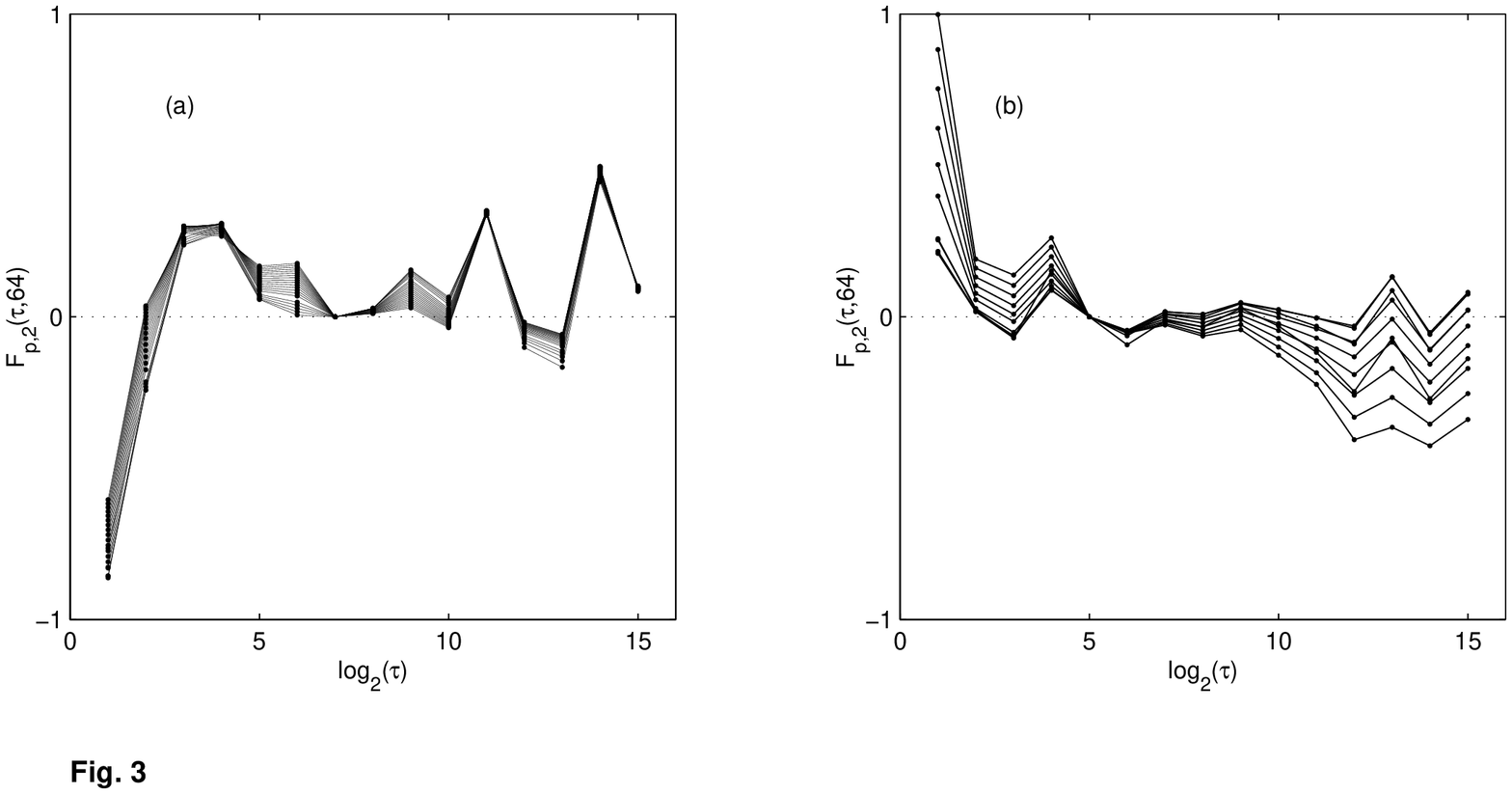}
\caption{Evidence of $\tau$-independent $S^\infty
(\tau)$. $F_{p,2}(\tau,64)$ vs. $\log_2(\tau)$ from (a) a 
healthy subject ($p=0.2\sim 5$) and (b) a CHF patient
($p=0.4\sim 2.6$). The $\beta$ needed in the calculation
of $\mu(p,q)$ (see \ref{9}) is obtained from that
estimated by (\ref{7}).}
\end{figure}

\begin{figure}
\centering
\includegraphics[width=.45\textwidth]{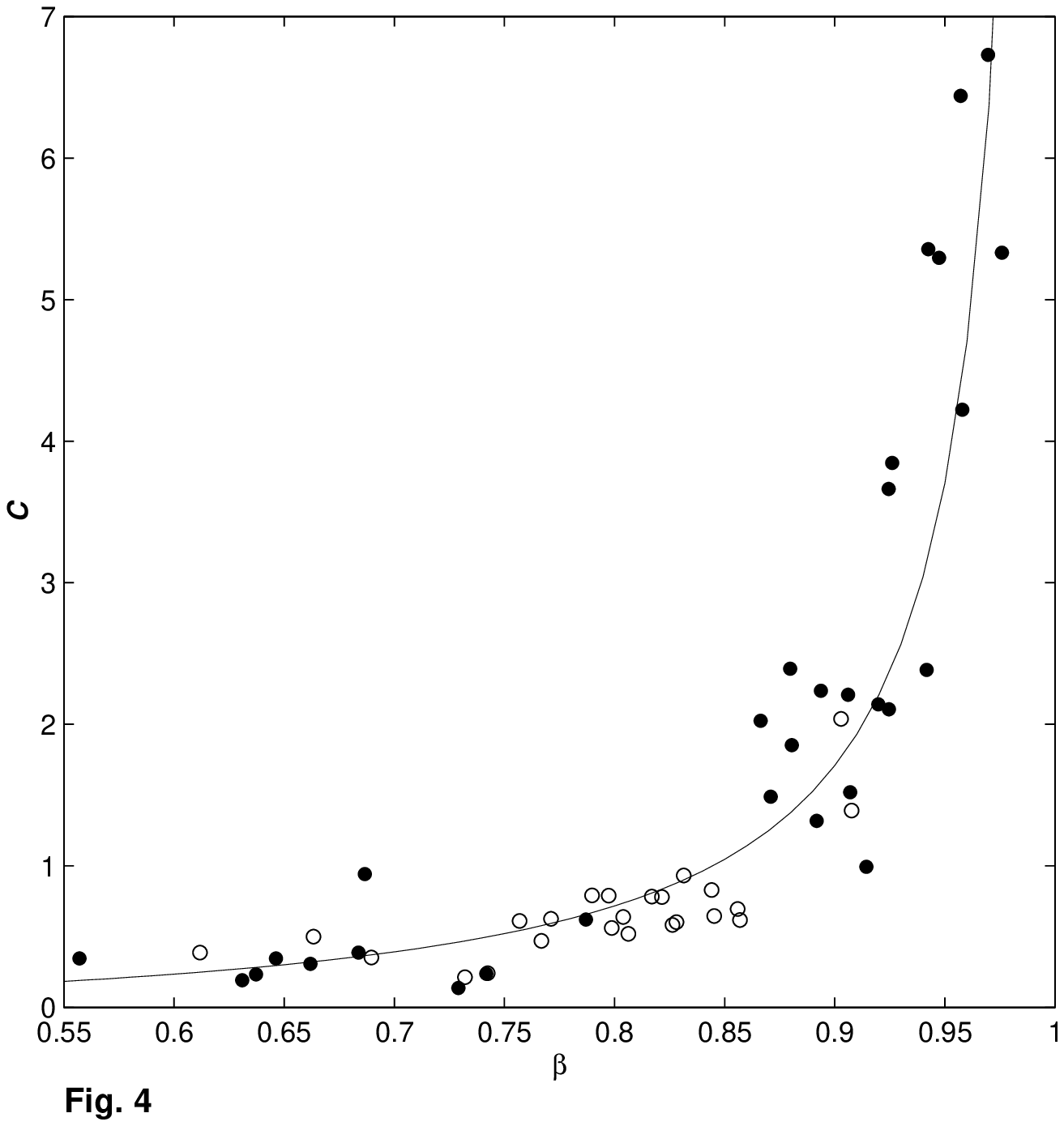}
\caption{$C$ vs. $\beta$ empirical law for healthy
subjects (open circles) and congestive heart failure
patients (solid circles). The solid line is the
fit $C = 0.2\beta/\ln(1/\beta)$.}
\end{figure}


\begin{thebibliography}{99} 
\bibitem{ref3} M. Kobayashi and T. Musha, IEEE Tans. Biomed. Eng. {\bf 
29}, 456 (1982). 
\bibitem{ref1}  
P.CH. Ivanov, L.A.N. Amaral, A.L. Goldberger, S. Havlin, M.G. 
Rosenblum, Z.R. Struzik and H.E. Stanley, Nature {\bf 399}, 
461 (1999). 
\bibitem{ref2}  
L.A.N. Amaral, P.Ch. Ivanov, N. Aoyagi, I. Hidaka, S. 
Tomono, A.L. Goldberger, H.E. Stanley, Y. Yamamoto, Phys. Rev. 
Lett. {\bf 86}, 6026 (2001). 
\bibitem{ref4} 
G.C. Butler, J. Floras, Clin. Sci. {\bf 92}, 545 (1997). 
\bibitem{ref5}  
D.C. Lin and R.L. Hughson, Phys. Rev. Lett. {\bf 86}, 
1650 (2001); D.C. Lin, Fractals {\bf 11}, 63 (2003). 
\bibitem{ref6}  
Z-S. She and E. Leveque, Phys. Rev. Lett. {\bf 72}, 336 (1994). 
\bibitem{ref7}  
E.S.C. Ching, Z-S. She, W. Su, Z. Zou, Phys. Rev. E 
{\bf 65}, 066303 (2002). 
\bibitem{ref8} 
R. Benzi, L. Biferale, S. Ciliberto, M. V. Struglia, 
and R. Tripiccione, Europhys. Lett. {\bf 32}, 709 (1995). 
\bibitem{ref9} R. Benzi, L. Biferale, S. Ciliberto, M. V. Struglia, 
and R. Tripiccione, Physica D {\bf 96}, 162 (1996). 
\bibitem{ref10}  
http://physionet.org. See also
A.L. Goldberger, L.A.N. Amaral, L. Glass, J.M. Hausdorff, 
P.Ch. Ivanov, R.G. Mark, J.E. Mietus, G.B. Moody, C.-K. Peng, and H.E. 
Stanley, {\it Circulation}, {\bf 101}, e215 (2000). 
\bibitem{ref11}  
Z-S. She and E.C. Waymire, Phys. Rev. Lett. {\bf 74}, 262 (1995). 
\end{thebibliography}
\end{document}